\documentclass[9pt]{project}
\usepackage{fullpage}
\usepackage[margin=1in]{geometry}
\usepackage{amsmath,xcolor}
\usepackage{url}
\usepackage{hyperref}
\usepackage{graphicx}

\makeatletter
\g@addto@macro{\UrlBreaks}{\UrlOrds}
\makeatother

\newenvironment{packeditemize}{
\begin{list}{\tiny$\bullet$}{
\setlength{\itemsep}{1.5pt}
\setlength{\labelwidth}{8pt}
\setlength{\leftmargin}{10pt}
\setlength{\labelsep}{3pt}
\setlength{\listparindent}{\parindent}
\setlength{\parsep}{1.5pt}
\setlength{\parskip}{1.5pt}
\setlength{\topsep}{1.5pt}}}{\end{list}}

\date{}

\title{\Large AppIntent: Intuitive Automation Specification Framework for Mobile App Testing} 
\author{Poornima Gopi}

\begin{document}\sloppy

\maketitle

\thispagestyle{empty}

\begin{abstract}
\hspace{\parindent}
The proliferation of mobile apps and reduced time in mobile app releases mandates the need for faster and efficient testing of mobile apps, their GUI and functional capabilities. Though, there are wide variety of open source tools and frameworks that are developed to provide automated test infrastructure for testing mobile apps. Each of these automation tools supports different scripting languages for automating the app testing. These frameworks fundamentally lacks the ability to directly capture the intent of the users who intend to effectively test the mobile app and its cross-app functional capabilities and performance without worrying about the low-level scripting language associated with each tool. 

Hence, to address this limitation, we propose a high-level intent-based automation specification language and APIs that could effectively address following aspects: ($i$) capture the test automation steps to be captured as high-level intents 
using intuitive automation specification language, and ($ii$) provides framework support to effectively capture the user’s behavior patterns effectively for testing the apps. We develop, {\em AppIntent} a high level automation specification language-based framework that directly captures the test automation intents with in and across multiple apps using high-level and intuitive language without worrying about {\em how} to actually develop scripts for automation. 


\end{abstract}

\section{Introduction}
\hspace{\parindent}
Today the mobile app functionalities are becoming quite complex with large set of functional capabilities and cross app communication events. Currently, there are more number of mobile web app users compared to desktop users~\cite{moreappsthan_desktop}. Also, these apps are developed as native, web and hybrid apps, which makes automation much more complex. Such heterogeneity in the application types, complexity in functional capabilities and cross app functionalities makes testing and automation much challenging. Also for testing the android platform with multiple mobile apps running at the same instance of time, which involves cross-app interaction and cross-app GUI events manually generated by human requires app testing engineers to effectively develop automation scripts. 
Today a wide varieties of open source and commercial tools are available for automating the app testing~\cite{bestapps_fortesting}. Currently, these automation tools adapts following techniques for generating the events for automating app testing: ($a$) {\em Fuzzy test}: Automation frameworks such as Monkey~\cite{monkey,monkeyrunner}, are developed to randomly generate events to fuzzy test the functionalities. But, the randomly generated events neither provides good code coverage nor provides effectively app testing considering the actual user behavior in lesser iterations of automation. ($b$) {\em Biased Randomness}: To avoid such complete randomness in event generation to app, newly developed frameworks effectively monitors the existing event and respectively generate next set of events with the help of human inputs i.e., providing biased randomness~\cite{dynodroid}, which helps to provide enhanced code coverage in shorter test cycles compared to complete randomness in event generation.


For automating the test scenarios, currently users need to either write code/scripts (i.e., using the framework supplied scripting languages such as ruby, python, java, robotium and so on) or use record-and-replay approaches to the test different scenario. Both these approaches results in increased testing and automation times if minor changes are required to the test scenario (i.e., requires rewrite of the automation code or re-recording the test scenario once again). In addition, if multiple apps are involved in testing with cross-layer communication across apps the automation becomes much more challenging. 

Hence, to make testing with automation frameworks much simpler, we proposed intuitive high-level automation specification language that provides following key benefits to its users: \\
\hspace{\parindent}
($i$) First, it allows test engineer or user to automate the testing by specifying the high-level intents using intuitive language such as described below (with out worrying about intricate scripting details).

\vspace{6 pt}
\noindent {\tt Automation specification syntax}: 

\noindent app-name\{cnn\} $\,\to\,$ login(username, credential) $\,\to\,$ checkpage\{tab1$\,\to\,$\ page1\}.stay-onpage\{5sec\} $\,\to\,$ logout().\\

\vspace{5 pt}
Each literal, function and $\,\to\,$ specifies the action to be carried out or event to be generated by the automation tool. Abstracting the low level language specifications and providing such high-level language allows the automation to be carried out faster and intent to be captured effectively.

($ii$) Second, it does not require user to record-and-replay for automating the mobile app test cases.

This is achieved by supplying the capabilities/features of mobile app available to our automation specification language as list of features, which user can choose to supply it to automation specification language. 

Our key contributions includes:
\begin{itemize}

\item
Develop intuitive automation specification language that allows complex automation steps to be specified in few lines compared to hundreds of lines coded as scripts.

\item
Allows users to easily customize automation test scenarios by simply tuning the parameters provided by the specification language specific to each functionality instead of changing the scripts or rerecording for replaying.

\item
Evaluate capability of our automation specification language framework with multiple apps which involves cross-app communication and complex capabilities of each app.

\end{itemize}



\subsection{Tool: Appium Studio} 
We are using Appium Studio~\cite{appium_studio} for our project. The main reason to use this tool is importantly this framework provides testing capabilities of web, native and hybrid apps, which we intend to evaluate our framework with. Also, multiple scripting language support provided by this studio framework will be abstracted with our common automation specification language illustrating that our framework can seamlessly work with multiple scripting languages or underlying techniques by supporting necessary APIs for translating our automation specification language to low level scripting language. Also the tool generates the results for the executed testcases and saves as images, which is very useful for test engineer to identify the failure so quickly without actually going through and studying the complete log file. This actually drastically reduces the debugging time of the test engineer. 

\subsection{Mobile Apps for Evaluation} \hspace{\parindent}
We plan to test tens of app which involves complex functional capabilities and cross-app functional events. For example, the web browser app (e.g., firefox), a social/new app (e.g., cnn) and video app (e.g., youtube). The cross-app communication between the web browser app and news channel apps with the video apps such as youtube will be tested with our framework.

\section{Background \& motivation}
\hspace{\parindent}
The main motivation for  developing a Intuitive automation specification language based framework to test the mobile applications was the increasing number of apps in the market and the problems that a app developer is facing in testing the app.

According to gartner report the global sales of smartphones to end users totaled 383 million units in the third quarter of 2017\cite{gartner}, with the increase in smartphones, the app usage has also increased drastically, from statistics  there are around 197 billion app downloads in 2017\cite{Appstatistics}. 
At present there are lot of free app construction frameworks available in the market and have attracted a large number of developers and organizations to develop and market their apps\cite{apptestingcompanies}. So the question is are the apps adequately tested before they are released into the app stores. To address this question there was a study done by Kochhar and et.al., has made a survey \cite{survey} to know the test automation culture of the app developers. They have studied apps and app developers from Fdroid and github and has clearly proved that developers are facing a lot of challenges in testing the apps. Some of the challenges they were facing were tools are cumbersome, poor documentation, a very steep learning curve, and compatibility issues.

To make the developer life easy in testing the app and to have a reliable app we have implemented a automation test framework and defining and proposing a new language which is very simple to  understand and implement that works on appium studio to test the app.

The more details of the framework and the specification language is explained in the system overview section.

\section{System Overview}

\hspace{\parindent}In this section, we describe the architectural details of {\tt AppIntent} automation framework. 
To address the challenges involved in automating the apps using the existing tool, we propose and implement an intent-based automation specification framework. {\tt AppIntent} abstracts the low-level automation details of the automation tools drastically reducing the learning time for each test engineer to learn the automation tool and perform the testing. In this architecture, we provide the ability to directly capture the intents of test engineer without worrying about the automation tool. For this we propose and implement an automation specification language which captures the ability automation intents much similar to the whiteboard drawing that the automation engineer uses for specifying the automation needs. The example automating different apps (i.e., cnn, torch, amazon, and weather app) using simple specification language is described below in Figure \ref{fig:sample_intents}.


\begin{figure}[h!]
	\centering
\includegraphics{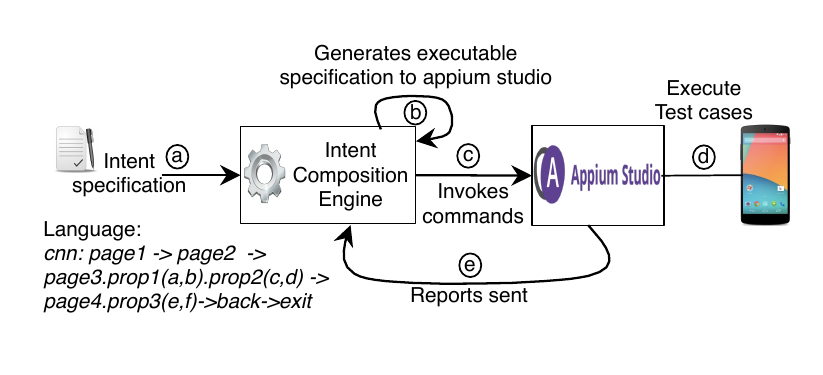}
\caption{High Level {\tt AppIntent} System Architecture}\label{fig:highleve_arch}
\end{figure}

\begin{table}[h!]
\footnotesize
\begin{tabular}{p{0.5in}|p{1.00in}|p{1.26in}} \hline
\textbf{Type} & \textbf{Symbol} & \textbf{Definition}  \\ \hline
Property specification keywords & credentials.\{\}, search.\{\}, select.\{\}, sleep.\{\}etc., & Keywords for capturing the properties of the App pages. \\ \hline

Action attributes / Keywords & App-commands-action, $\rightarrow$ (Next step) \{\} commands inside braces are key,value pairs, ".`` defines the condition to be specified on each page & Attributes or keywords used to specify the actions to be taken in the app. \\ \hline
\end{tabular}
\caption{Intent language definition}\label{}
\end{table}

\vspace{5 pt}
As shown in the Figure \ref{fig:highleve_arch}, the intents specified by admin using the intent specification language is digested by the {\em intent composition engine}. The {\em composition engine} then translates the specifications captured in a input file, translated into executable automation script that can be executed directly on the {\tt Appium} studio. Appium studio executes the respective test cases that are captured in the executable automation script on to the mobile app. The results of the automation script execution will be sent back to the intent composition engine as report.


\subsection{Intent Specification}

\hspace{\parindent}The automation intent specification language is has following fundamental skeleton for capturing the automation intents. The initial key in the intent represents the app name. From the example described in Figure \ref{fig:sample_intents} the {\tt cnn}, {\tt flashlight} and {\tt weather} represents the name of the app on which the administrator's automation intent need to be executed. The next names separated by $\rightarrow$ each represents the page on which the operation is being carried out inside the app. For example, in the intent {\tt I1}, we will go to the pages named {\tt money} followed by {\tt share} tab, then {\tt watchnow}, {\tt back} page and finally {\tt exit} the app. We capture the operations to be carried out onto each page using the .dot operator. For example. once reaching the login page, we can carryout authentication and then move to balance page as follows: {\tt login}.authentication({\em userid},{\em password}) $\rightarrow$ {\tt balance}. Similar operations like scroll(down), stay-on-page($sec$) operations are some such example of operations that can be captured in our specification language. 

List of example intents supplied for testing apps are listed below in the Figure \ref{fig:sample_intents}. {\tt I1} represents the intent to test the {\tt cnn} app. Similarly, {\tt I2} and {\tt I3} for testing the flashlight and weather apps.

\vspace{10pt}
\noindent {\tt I1: cnn}: $topnews \rightarrow u.s.politics \rightarrow money \rightarrow share \rightarrow watchnow \rightarrow back \rightarrow exit$

\vspace{2pt}
\noindent {\tt I2: flashlight}:  $settings \rightarrow smartcharge \rightarrow switch \rightarrow back \rightarrow home \rightarrow exit$

\vspace{2pt}
\noindent {\tt I3: weather}: $image \rightarrow settings \rightarrow aboutapp \rightarrow back \rightarrow units \rightarrow home \rightarrow exit$

\vspace{2pt}
\noindent {\tt I4: Amazon}: $launch \rightarrow sleep \rightarrow signin \rightarrow existinglogin \rightarrow   enterdetails \rightarrow credentials.\{username:mobiletestaa@gmail.com\} \rightarrow continue \rightarrow passwordfield \rightarrow credentials.\{password:testingapp\} \rightarrow logincreate \rightarrow search.\{searchbar:kidsbooks\} \rightarrow enter \rightarrow select.\{itemnumber:1\} \rightarrow clickcart \rightarrow checkout
\rightarrow exit$

\begin{figure}[h!]
\vspace{-0.10in}
\caption{Sample intents for automating three different apps are described. Here each word represents the page to be loaded from the mobile app.}\label{fig:sample_intents}
\end{figure}

\subsection{Intent Mappings} 

\hspace{\parindent}
The mapping used for building the executable commands for running the cnn app testing using {\tt Appium} studio is listed below in Figure \ref{fig:sample_mappings}. The mappings plays key role in translating the intent key work into the executable command that can be executed on to the Appium studio. For example, press of {\tt WatchNow} button on the page is captured using the syntax in the mapping file, which will be used by our composition engine to generate the executable script using the mappings. 

\begin{figure}[h!]
\centering
\includegraphics{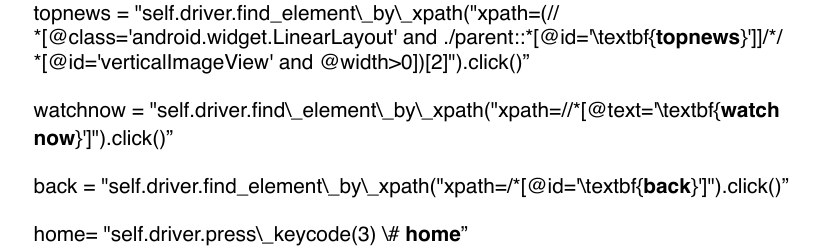}
\caption{Mappings used in generating the {\tt Appium} studio specific executable commands at stage (b) of Figure \ref{fig:highleve_arch}.}\label{fig:sample_mappings}
\end{figure}

\subsection{Limitations and next stage tasks}
\hspace{\parindent}
The limitations of the current version of the code includes: 

\begin{packeditemize}

\item [1.] We need to manually identify each of the objects or page names inside the app for specifying it as an {\tt intent} by administrator. Though the object structure of the app will be automatically extracted as the abstraction, which can then be used directly for specifying the automation intents.

\item [2.] We need to enhance the tool to capture the operations that need to be carried onto each page as a single word intent, which will be delivered in the next release.

\end{packeditemize}

\subsection{Future Work} 

\hspace{\parindent}
In this section, we describe about the limitations of our existing framework and potential for future enhancements to the tool and future research directions.\\

We strongly believe that this high-level automation specification language framework could result in drastically reducing the time required to automate the mobile apps testing and provide common framework to integrate different mobile app testing tools.
This work has potential for following future works.

\begin{itemize}

\item The abstraction and automation specification framework that we propose has potential to bring multiple test tools under same testbed environment allowing the test engineer to capture his/her automation intents irrespective of the test tool. 

\item Abstract the objects from the app pages automatically by getting the dorm structure of the app and identifying all the objects and pages in the app automatically.

\item Expand the specification language with various different scenarios of the apps to successfully handle the complete end to end test cases.

\end{itemize}


\section{Testbed} 
\begin{figure}[h!]
\includegraphics[height=2in,width=3.1in]{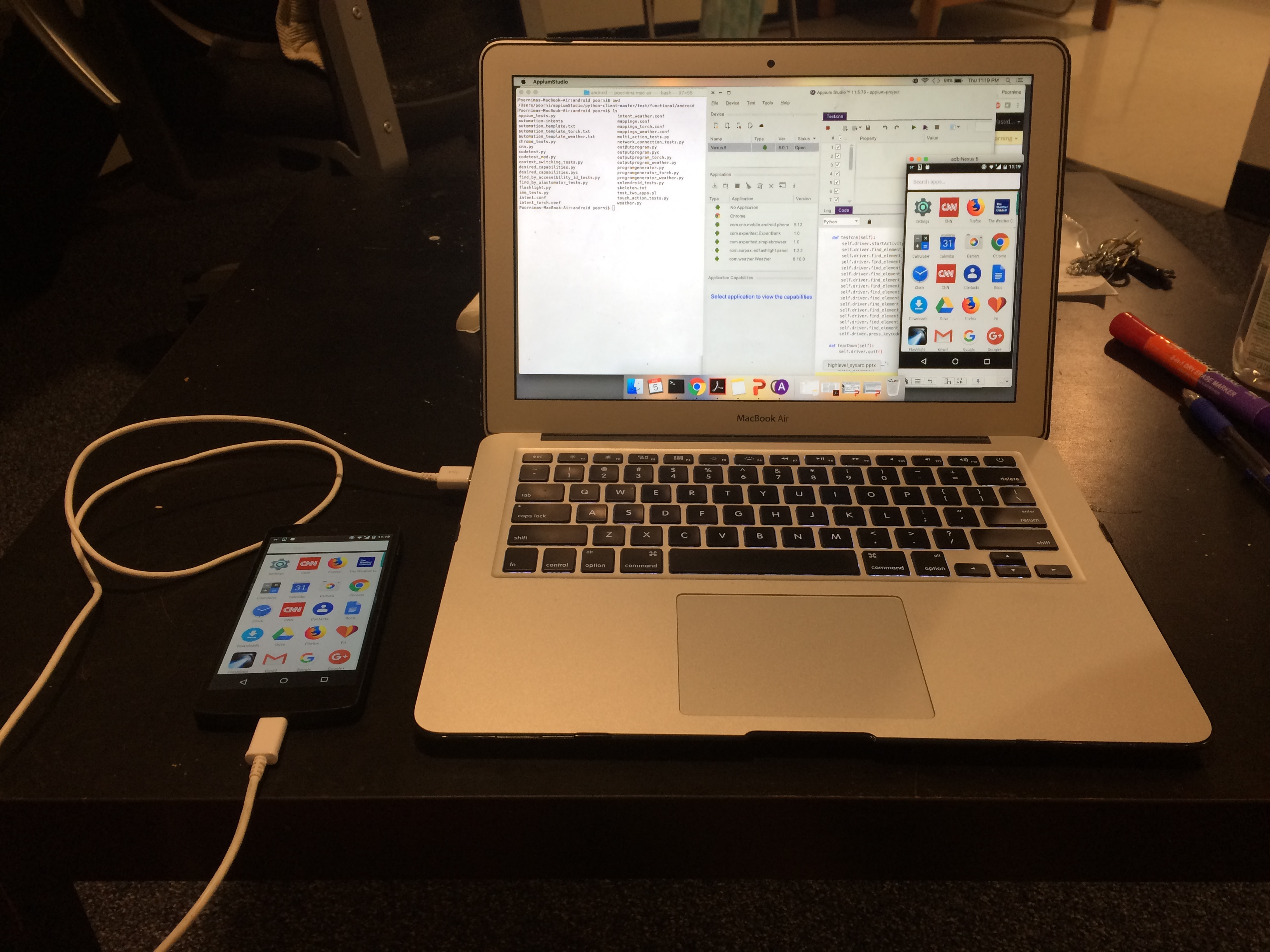}
\caption{Testbed for {\tt AppIntent}}\label{fig:highleve_arch}
\end{figure}
We implemented the {\em AppIntent} framework in python. We run the Appium studio on Macbook air Intel core 1.6Ghz, 4GB RAM. We tested the setup with Nexus 5 Android mobile attached to the laptop using the USB cable.

As of now we are having a basic test case template used to run a test case on the app. we are specifying the intent in a file with the intent specification language using the keywords as shown in figure2, and we are manually identifying the objects in the test app and maintaining the mappings i.e., identified objects as values of keywords into mappings file. We try to generate the code automatically using the intent and mappings key value pairs and saved to the template which automatically gives the code for the user intended test case. But we are planning to implement a graph based abstraction to automatically identify the objects of the app.

We have currently tested our utility with following eight apps: 
($i$) cnn (news), ($ii$) weather,  ($iii$) flashlight, ($iv$) raaga (music), ($v$) walmart (shopping) ($vi$) amazon (shopping), ($vii$) bbc (news) ($viii$) pacer (excercise). We will be testing more apps in future with extended language specifications. \\

\section{Evaluations}

Let us understand the step by step procedure involved in testing the app using this Intent language based specification framework. We can consider the app amazon(shopping), first we will specify the intents for testing the amazon app as below\\

{\tt \noindent {\tt Amazon}: launch $\rightarrow$ sleep $\rightarrow$ signin $\rightarrow$ existinglogin $\rightarrow$   enterdetails $\rightarrow$ credentials.\{username:mobiletestaa@gmail.com\} $\rightarrow$ continue $\rightarrow$ passwordfield $\rightarrow$ credentials.\{password:testingapp\} $\rightarrow$ logincreate $\rightarrow$ search.\{searchbar:kidsbooks\} $\rightarrow$ enter $\rightarrow$ select.\{itemnumber:1\} $\rightarrow$ clickcart $\rightarrow$ checkout
$\rightarrow$ exit\\}

The testcase equivalent to the above intent is launch the amazon app and wait for the login screen if there is existing account on amazon then use the username and password given inside the braces and login to the account. After login search for the kids books and select the first item present in the list, then add it to the cart and checkout. Once its done logout and exit the app.

The above intent is provided to the Intent composition engine, a mapping file for all the objects of the amazon app are manually added (we are planning to get the dorm structure of the app and capture the mappings automatically using graph based approach in the future work) and are provided to the composition engine. The mappings file of amazon has objects as shown below:\\

\noindent {\tt signin = "self.driver.find\_element\_by\_xpath\\("xpath=//*[@text='Hello. Sign In']").click()"\\
existinglogin = "self.driver.find\_element\_by\_xpath\\("xpath=//*[@text='Login. Already a customer? ' and @class='android.widget.RadioButton']").click()"\\
search = "self.driver.find\_element\_by\_xpath\\("xpath=//*[@id='rs\_search\_src\_text']").click()"\\}

The composition engine digests the intents and extracts the mappings associated with each intent and generates the appium executable code using the automation template file that is provided to the appiumstudio. The automation template file will have the details of the device i.e., the mobile phone that is used to test the app like device id, appium device driver location to import the dependency files and also We need to provide the appname and results location to the appiumstudio. The sample format of the automation template is shown below\\

{\tt \noindent import unittest\\
import time\\
from appium import webdriver
reportDirectory = 'reports'\\
reportFormat = 'xml'\\
testName = 'Untitled'\\
driver = None\\
def setUp(self):\\
self.dc['reportDirectory'] = self.reportDirectory \\
self.dc['reportFormat'] = self.reportFormat\\
self.dc['testName'] = self.testName\\
self.dc['udid'] = '03ab1c13003c0097'\\
self.dc['appPackage'] = 'com.amazon.mShop.android.\\shopping'\\
self.dc['appActivity'] = 'com.amazon.mShop.home.HomeActivity'\\
self.dc['platformName'] = 'android'\\
self.driver = webdriver.\\Remote('http://localhost:4723/wd/hub',self.dc)\\

}

The generated appium executable testcase is then sent to the tool and executed, which inturn is executed on the real device. Once the execution is completed the results are sent back to the composition engine and displayed both on the console and also saved as images in the results location that we have provided.

So we were able to successfully 

Once the test is executed the results of the execution can be seen on the commandline and looks like below:\\

\noindent \textbf{Poornimas-MacBook-Air:android poorni\$} {\tt python programgenerator\_amazon.py} \\

\noindent {\tt \# Creating outputprogram\_amazon.py file for generating the code\\

\noindent \# Intents are :\\
launch $\rightarrow$ sleep $\rightarrow$ signin $\rightarrow$ existinglogin $\rightarrow$ enterdetails $\rightarrow$ credentials.\{username:mobiletestaa@gmail.com\} $\rightarrow$ continue $\rightarrow$ passwordfield $\rightarrow$ credentials.\{password:testingapp\} $\rightarrow$ logincreate $\rightarrow$ search.\{searchbar:kidsbooks\} $\rightarrow$ enter $\rightarrow$ select.\{itemnumber:1\} $\rightarrow$ clickcart $\rightarrow$ checkout $\rightarrow$ exit\\
\textbf{Poornimas-MacBook-Air:android poorni\$} testUntitled (\_\_main\_\_.MultiActionTests) ...\\ 
ok\\
-----------------------------------\\-----------------------------------\\
Ran $1$ test in $121.386$s\\
OK\\}



\section{Related Work}
\hspace{\parindent}
Currently, these automation tools adapts following techniques for generating the events for automating app testing: ($a$) {\em Fuzzy test}: Automation frameworks such as Monkey~\cite{monkey,monkeyrunner}, are developed to randomly generate events to fuzzy test the functionalities. But, the randomly generated events neither provides good code coverage nor provides effectively app testing considering the actual user behavior in lesser iterations of automation. ($b$) {\em Biased Randomness}: To avoid such complete randomness in event generation to app, newly developed frameworks effectively monitors the existing event and respectively generate next set of events with the help of human inputs i.e., providing biased randomness~\cite{dynodroid}, which helps to provide enhanced code coverage in shorter test cycles compared to complete randomness in event generation.
Though there are many automation test frameworks available in the market, to my knowledge this is the first work to focus on the difficulties and problems faced by the app developers in testing the app and proposed a new framework which will be very simple and does not require any experience.

\section{Conclusion}
\hspace{\parindent}
Developed an intuitive automation specification language that allows complex automation steps to be specified and has tested around 8apps and has shown that the framework was capable of generating the testcases  without actually record and replay or by writing the code for the testcases. The specification language implemented in this framework if extended to full extent can definitely reduce the pain of the developers in testing the apps. This framework can be extended and can be integrated with more testing tools.
Showcase ability to capture the Intent and run the testcases on different kinds of apps. The advantages of our framework are  no programming skills are required, easy and fast to implement \& Customize test cases. No compatability issues with different platforms. Can be tested on both emulators as well as on the real devices without worrying about the underlying tool or coding language used.

\bibliographystyle{abbrv}
\begin{small}
\bibliography{references}
\end{small}

\end{document}